\documentclass[sigconf]{acmart}
\AtBeginDocument{%
  }

\setcopyright{acmlicensed}
\copyrightyear{2018}
\acmYear{2018}
\acmDOI{XXXXXXX.XXXXXXX}
\acmConference[Conference acronym 'XX]{Make sure to enter the correct
  conference title from your rights confirmation email}{June 03--05,
  2018}{Woodstock, NY}
\acmISBN{978-1-4503-XXXX-X/2018/06}

\usepackage{amsmath}
\usepackage{multirow}
\begin{document}

\title{Spacetime-GR: A Spacetime-Aware Generative Model for Large Scale Online POI Recommendation}


\author{Haitao Lin}
\email{lht416932@alibaba-inc.com}
\affiliation{
  \institution{AMAP, Alibaba Group}
  \city{Beijing}
  \country{China}
}

\author{Zhen Yang}
\authornote{Corresponding Author}
\email{zhongming.yz@alibaba-inc.com}
\affiliation{
  \institution{AMAP, Alibaba Group}
  \city{Beijing}
  \country{China}
}

\author{Jiawei Xue}
\email{xuejiawei.xjw@alibaba-inc.com}
\affiliation{
  \institution{AMAP, Alibaba Group}
  \city{Beijing}
  \country{China}
}

\author{Ziji Zhang}
\email{zhangziji.zzj@alibaba-inc.com}
\affiliation{
  \institution{AMAP, Alibaba Group}
  \city{Beijing}
  \country{China}
}

\author{Luzhu Wang}
\email{luzhu.wlz@autonavi.com}
\affiliation{
  \institution{AMAP, Alibaba Group}
  \city{Beijing}
  \country{China}
}

\author{Yikun Gu}
\email{yikun.gu@alibaba-inc.com}
\affiliation{
  \institution{AMAP, Alibaba Group}
  \city{Beijing}
  \country{China}
}

\author{Yao Xu}
\email{xuenuo.xy@alibaba-inc.com}
\affiliation{
  \institution{AMAP, Alibaba Group}
  \city{Beijing}
  \country{China}
}

\author{Xin Li}
\email{beilai.bl@alibaba-inc.com}
\affiliation{
  \institution{AMAP, Alibaba Group}
  \city{Beijing}
  \country{China}
}

\begin{abstract}
    Building upon the strong sequence modeling capability, Generative Recommendation (GR) has gradually assumed a dominant position in the application of recommendation tasks (\textit{e.g.}, video and product recommendation). However, the application of Generative Recommendation in Point-of-Interest (POI) recommendation—where user preferences are significantly affected by spatiotemporal variations—remains a challenging open problem. In this paper, we propose Spacetime-GR, the first spacetime-aware generative model for large-scale online POI recommendation. It extends the strong sequence modeling ability of generative models by incorporating flexible spatiotemporal information encoding. Specifically, we first introduce a geographic-aware hierarchical POI indexing strategy to address the challenge of large vocabulary modeling. Subsequently, a novel spatiotemporal encoding module is introduced to seamlessly incorporate spatiotemporal context into user action sequences, thereby enhancing the model's sensitivity to spatiotemporal variations. Furthermore, we incorporate multimodal POI embeddings to enrich the semantic understanding of each POI. Finally, to facilitate practical deployment, we develop a set of post-training adaptation strategies after sufficient pre-training on action sequences. These strategies enable Spacetime-GR to generate outputs in multiple formats (\textit{i.e.}, embeddings, ranking scores and POI candidates) and support a wide range of downstream application scenarios (\textit{i.e.}, ranking and end-to-end recommendation). We evaluate the proposed model on both public benchmark datasets and large-scale industrial datasets, demonstrating its superior performance over existing methods in terms of POI recommendation accuracy and ranking quality. Furthermore, the model has been successfully deployed in online POI recommendation services that scale to hundreds of millions of POIs and users. To the best of our knowledge, it is the first generative model successfully applied to large-scale industrial online POI recommendation systems.
\end{abstract}



\maketitle

\section{Introduction}
In recent years, Generative Recommendation \cite{yang2025gr} has emerged as the dominant paradigm in recommender systems. Traditional recommendation methods \cite{covington2016deep,wang2011cascade,zhou2018deep} typically follow a three-stage pipeline: recall \cite{covington2016deep,li2019multi}, pre-ranking \cite{ma2021towards,wang2020cold}, and ranking \cite{10.5555/3172077.3172127,cao2022sampling,pi2020search}. In contrast, generative methods \cite{kang2018self,rajput2023recommender} employ a unified model to directly generate recommendation results, often utilizing LLMs as their foundational architecture \cite{jiang2025large, lin2024data}, thereby outperforming traditional methods.

\begin{figure}[htbp] 
  \centering
  \includegraphics[width=0.45\textwidth]{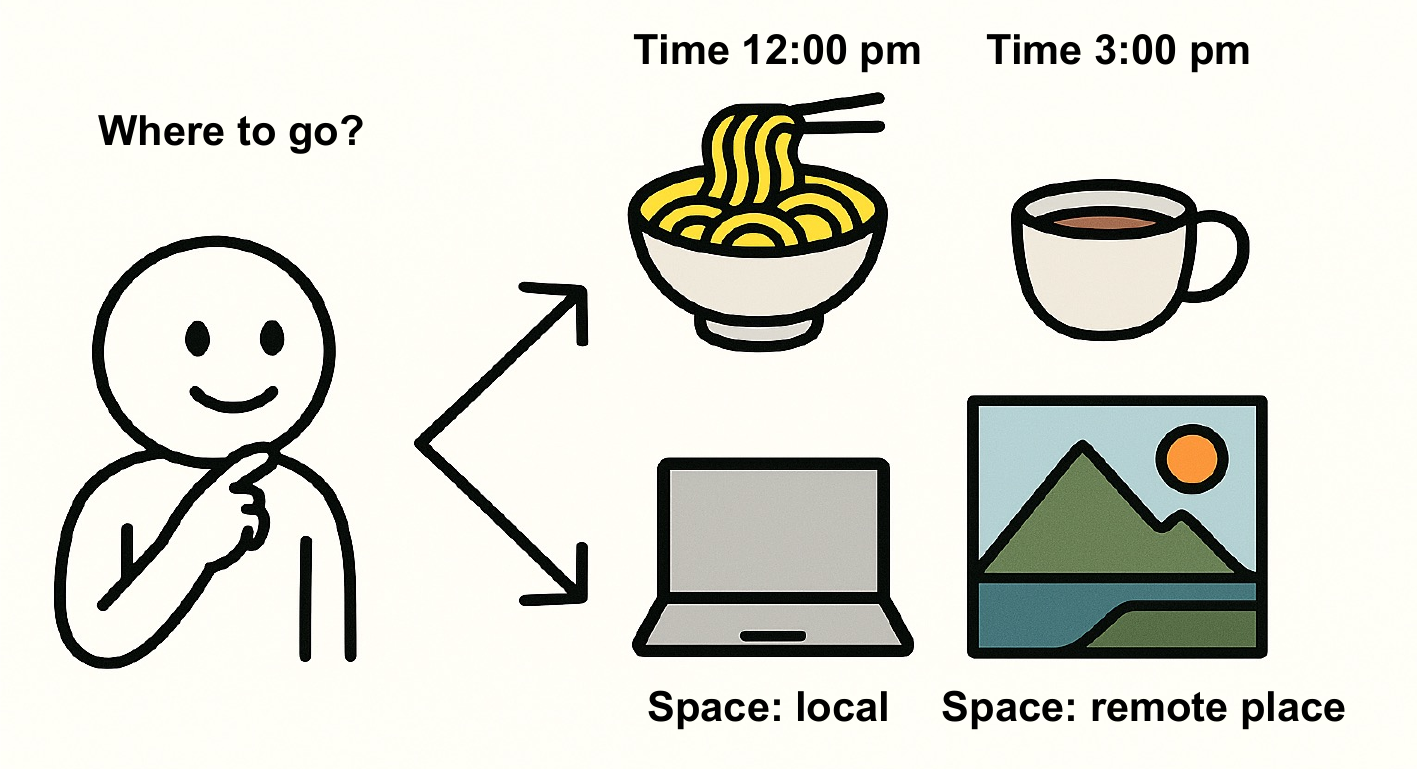}
  \caption{\label{fig: Intro} An example illustrating the spatiotemporal sensitivity of POI recommendation. From the time perspective, users are more likely to visit restaurants at noon but may prefer cafes around 3 PM. From the space perspective, users tend to show work-related interests in their local areas, whereas they prefer to visit attractions when in remote locations.}
\end{figure}

However, in the field of Points of Interest (POI) recommendation, generative approaches remain relatively underexplored. POI recommendation involves predicting POIs that a user might be interested in based on their historical action trajectories. Previous works on POI recommendation mainly experiment on small-scale public datasets \cite{li2024large,wu2020personalized,zhang2021interactive}. In the context of large-scale practical POI recommendation, several challenges arise: (1) \textbf{Modeling Large Vocabulary}: Directly modeling representations for hundreds of millions of POIs leads to an excessively large vocabulary, making the training process computationally prohibitive. (2) \textbf{Learning Spatiotemporal Sensitivity}: Unlike item or video recommendation tasks, user action sequences in POI recommendation include not only the attributes of POIs but also associated spatial and temporal information. As illustrated in Figure \ref{fig: Intro}, a user's interests may vary significantly under different spatiotemporal conditions. Therefore, effectively modeling the spatiotemporal context becomes a critical aspect of this task. (3) \textbf{Encoding sufficient POI information}: As the number of POIs grows up, those in user sequences typically exhibit a long-tail distribution, with most appearing infrequently. Consequently, models struggle to encode these rare POIs effectively.

To address the critical bottlenecks mentioned above, we propose Spacetime-GR, a spacetime-aware generative model for large scale online POI recommendation. Spacetime-GR extends the sequence
modeling ability of generative models by incorporating flexible spatiotemporal information encoding methods. First, a \textbf{geographic-aware hierarchical POI indexing strategy} is introduced to address the large vocabulary problem and enhance the spatial information in POI indexes. Second, we introduce a novel \textbf{spatiotemporal encoding module} that integrates user spatiotemporal information into the sequence, thereby enhancing the spatiotemporal sensitivity of the recommendation results. Third, \textbf{LLM-encoded multimodal POI embeddings} are added to Spacetime-GR for improving the model's understanding of POI semantics. Moreover, we introduce a two-stage training paradigm tailored for POI recommendation applications. The first stage is pre-training, where we learn latent patterns in user actions and the relationships between POIs by predicting the next token in large-scale user action sequences. A specialized data filtering strategy and loss function are adopted to capture users' real interests in POIs. The second stage is post-training, where the model is fine-tuned using recommendation data to meet specific application requirements. We propose three post-training strategies: \textbf{Embedding-based Ranking SFT} generates embeddings for users and POIs, serving as additional features for downstream ranking modules; \textbf{Generative Ranking SFT} outputs scores for POIs, which can be directly used for ranking; \textbf{DPO Alignment} provides end-to-end POI recommendation results, enhancing the model's joint recall and ranking ability.


The key innovations of this paper are as follows:

(1) We propose Spacetime-GR, a spacetime-aware generative model for online POI recommendation. Three critical optimizations, including spatiotemporal encoding module, geo-aware hierarchical POI indexing strategy, and LLM-encoded multimodal POI embeddings are introduced to address the challenges in industrial POI recommendation.

(2) We design an effective training and application framework for POI recommendation. Leveraging pre-training and fine-tuning methods, Spacetime-GR could provide various types of outputs supporting for different downstream scenarios.

(3) We are the first to successfully apply the generative model to industrial POI recommendation systems, catering to recommendation services involving hundreds of millions of users and POIs. 

\section{Related Work}

\subsection{Sequential Recommendation}

Based on the model structure, common approaches can be categorized into two types: discriminative and generative.


\textbf{Discriminative methods} mainly adopt an encoder-based network to rank candidate items based on user sequences. Traditional methods use neural networks for sequence modeling and score prediction \cite{pi2020search, zhou2018deep}. With the advancement of pre-trained models \cite{wu2024survey,zhao2023survey}, many works have begun to decouple the sequence modeling component from the discriminative scoring model. For instance, LEARN \cite{jia2024knowledge} independently pre-trains the representation of users and items based on users' behaviors. Similarly, HLLM \cite{chen2024hllm} leverages the text encoding capability of language models to hierarchically encode user behavior sequences: the lower-level LLM encodes item information, while the upper-level LLM encodes user information. These representation features are then integrated into the ranking model to enhance its Click-Through Rate (CTR) estimation ability.

\textbf{Generative methods} employ a decoder-based architecture to directly generate recommendation results in an end-to-end manner. This paradigm involves encoding user-interacted items using unique identifiers (IDs), and predict the probability of each ID, from which the top-K IDs with the highest probabilities are selected as the output results. SASRec \cite{kang2018self} predicts the next user-interacted item through an autoregressive approach. It utilizes a single token to represent an item, which can lead to training difficulties when dealing with a large number of items. TIGER \cite{rajput2023recommender} incorporates the idea of RQ-VAE \cite{zeghidour2021soundstream} to learn to transform items into multiple semantic IDs, largely reducing the vocabulary size. Building upon semantic ID encoding, OneRec \cite{deng2025onerec} employs a Mixture of Experts (MoE) architecture \cite{dai2024deepseekmoe} and a Direct Preference Optimization (DPO) strategy \cite{rafailov2023direct} to further improve the recommendation ability. \cite{tan2024idgenrec} adopts LLM to generate more semantically rich indexes. \cite{zhai2024actionsspeaklouderwords} proposes HSTU, a new transformer framework for better modeling sequences, which is followed by \cite{huang2025largescalegenerativeranking} on ranking tasks. Moreover, some recent works \cite{10.1145/3604915.3608857,10.1145/3627673.3679611,10.1145/3701551.3703496,chen2025dlcrec,geng2022recommendation,jiang2025largelanguagemodeluniversal} design specialized prompts and directly finetune LLMs to predict the recommended items, fully leveraging the inference ability of the model while increasing the sequence length and the computational cost as well. 


In contrast to the aforementioned methods, our study focuses on addressing the challenge of POI recommendation in spatiotemporal contexts. Although we adopt the generative approach for training, we provide both discriminative and generative approaches for application of Spacetime-GR, demonstrating that both can enhance POI recommendation performance in real-world scenarios.

\subsection{POI Recommendation}

\textbf{POI Recommendation} is a sequential recommendation task specifically designed for POI contexts. The input consists of a user's historical POI visitation information, while the output is the prediction of the user's next visited POI. Common approaches include methods based on Recurrent Neural Networks \cite{feng2018deepmove,huang2019attention,lian2020geography,zhao2020go}, Memory Networks \cite{zhou2019topic}, Generative Adversarial Networks \cite{zhou2019adversarial}, Transformers \cite{yan2023spatio}, Graph Neural Networks \cite{lim2020stp} and Large Language Models \cite{jiawei2024large, li2024large, liu2024nextlocllm,wang2023would}. Most methods have been primarily evaluated on small-scale datasets, focusing on modeling offline check-in behaviors. In contrast, \cite{chen2021curriculum} expands the scale to millions of POIs, focusing mainly on transfer learning for POI prediction across different cities. 

Our work differs from existing research in two aspects. First, we focus on online POI recommendation, where the target is to recommend POIs that users are more likely to click rather than arrival. Second, we focus on large-scale application scenarios involving over 100 million POIs and users. Within this context, directly fine-tuning LLMs becomes impractical due to the long sequence input and associated computational costs. To address these challenges, we propose the spacetime-GR, which balances the powerful capability of generative models with practical online applicability.

\section{Methods}

\subsection{Task Definition}

First, we define the spacetime-aware online POI recommendation task as follows: 

Given dataset $D$ where each sample $S$ is a user's action sequence $\{s_1, s_2, ..., s_n\}$. Each item in $S$ is a user action, representing a user's online interaction with POI. It contains timestamp information $t$, user geographic information $g^u$, POI index $p$, POI geographic information $g^p$, POI category information $c$, and action type $a$. Besides, each sequence $S$ is provided with the user's profile information $up$. An example of the user action is presented in Table \ref{tab: Data Illustration}. In the recommendation task, the input is the user profile information $up$, user history action sequence $\{s_1, s_2, ..., s_m\}$, the timestamp of the $(m + 1)\text{th}$ action $t_{m+1}$, the user geographic information of the $(m + 1)\text{th}$ action $g^u_{m+1}$, and the output is the POI index of the $(m + 1)\text{th}$ action $p_{m+1}$.

\begin{table}[htbp]
\caption{\label{tab: Data Illustration} \normalsize Data format of an action in the spacetime-aware POI recommendation task.}
\centering
\resizebox{0.47\textwidth}{!}{
\begin{tabular}{l|l|l}
\toprule[1.0pt]
Key   &  Explanation  & Example Value   \\ 
\addlinespace[1pt]
\hline
\addlinespace[2.0pt]
time $t$ & a 13-digit timestamp & 1709805845148 \\
user geo info $g^u$ & the longitude and latitude of user location & x: 118.2252, y: 24.6001 \\
POI $p$ & the index of POI & 123 \\
POI geo info $g^p$ & the longitude and latitude of POI location & x: 118.3468, y: 24.1159 \\
POI category $c$ & the type of POI & food, French food \\
action type $a$ & the type of action & click \\
\addlinespace[1pt]
\bottomrule[1.0pt]
\end{tabular}
}
\end{table}

\subsection{Spacetime-GR}

\begin{figure*}[htbp] 
  \centering
  \includegraphics[width=\textwidth]{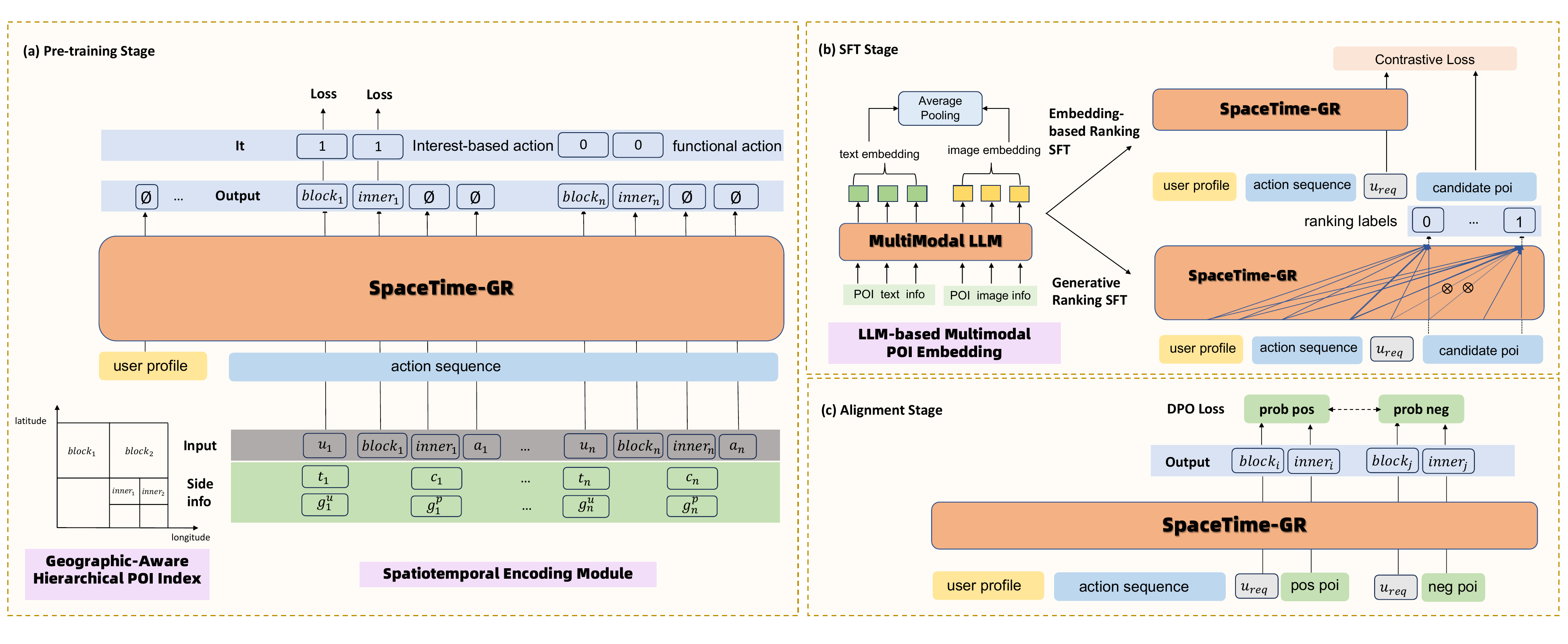}
  \caption{\label{fig: Model} The overall structure of Spacetime-GR, including three stages (pre-training, SFT, alignment).}
\end{figure*}

\subsubsection{Pretraining Stage}

\label{subsection: Pretraining Stage}
As shown in Figure \ref{fig: Model} (a), similar to the training paradigm of LLMs, the initial stage of our framework involves pre-training on users' action data. We utilize one year's action data\footnote{All the data are anonymous.} from sampled users on Amap\footnote{www.amap.com}, one of the most popular map service application in China, and organize it into sequential samples in chronological order, where each sample corresponds to a user's action sequence. 

\paragraph{\textbf{Data Cleansing}}
In the text pre-training process, data cleansing is a crucial step that aids in removing low-quality content from the collected data, thereby preventing the model from being negatively influenced by noisy or biased samples \cite{zhao2023survey}. Similarly, we conduct data cleansing at both action and sequence levels.

\subparagraph{\textbf{Action Level.}} User actions on map applications can be categorized into two modes: functional and interest-based actions. Functional actions, such as navigating to home or searching for a hospital, are initiated by users to fulfill specific purposes. Interest-based actions, on the other hand, involve actions like clicking on POI related to dining and entertainment after work. Compared with functional actions, interest-based actions are more informative for recommendation, as they help infer users' potential interests in the absence of explicit intent. Therefore, we classify actions at the POI level based on their intent. Specifically, we consider both the POI category and the ratio of active searches to passive clicks associated with the POI. Based on these factors, each action $s_i$ is labeled as either functional ($It_i=0$) or interest-based ($It_i=1$). During training, we focus solely on predicting actions where $It_i=1$.

\subparagraph{\textbf{Sequence Level.}} User actions on map applications vary significantly. While some users exhibit a ``two-point'' linear mode between home and office, others engage in diverse activities such as traveling, dining, and shopping. Compared to the former, the latter with various actions are more valuable. To address this, we design a metric named richness to filter sequences, defined as:

$$R = \frac{\text{The number of different POIs}}{\text{The number of actions}}$$

Sequences with more repeated POIs will have lower richness scores. Ultimately, we discard sequences with low richness scores ($R < 0.3$) and use the remaining samples for training.

The data filtering results are presented in Table \ref{tab: Data Cleansing}. We first apply filtering at the action level, followed by sequence-level filtering. Filtering out data with low quality helps prevent the model from converging to simplistic rule-based behaviors, encouraging it to focus more on uncovering users' deeper interest needs. The whole data cleansing stage enables faster convergence and allows the model to achieve better performance in the pre-training stage.

\begin{table}[htbp]
\caption{\label{tab: Data Cleansing} \normalsize Data cleansing statistic information (B stands for Billion).}
\centering
\resizebox{0.47\textwidth}{!}{
\begin{tabular}{l|ccc}
\toprule[1.0pt]
Filtering Step & Sequence Num & Action Num & Filter Ratio (action level) \\ 
\addlinespace[1pt]
\hline
\addlinespace[2.0pt]
coarse data & 1.06B & 77B & - \\
+ action level & 0.95B & 31.7B & 58.2\% \\
+ sequence level & 0.62B & 28.8B & 9.1\% \\
\addlinespace[1pt]
\bottomrule[1.0pt]
\end{tabular}}
\end{table}

\paragraph{\textbf{Model Structure}} After obtaining the pre-training data, we construct our pre-training model as depicted in Figure \ref{fig: Model} (a). The entire framework follows a decoder-only transformer architecture, based on the Llama 2 framework \cite{touvron2023llama}, with specific optimizations tailored to our task.

In the input part, to efficiently model the ID information of hundreds of millions of POIs, we propose a \textbf{geographic-aware hierarchical POI indexing strategy}. Inspired by the concept of semantic ID \cite{rajput2023recommender}, we use two tokens to represent each POI $p_i$
. The first token, $block_i$, indicates the block where $p_i$ is located (within a 5km by 5km area). Within a block containing $K$ POIs, we assign indexes ranging from 1 to K to represent these POIs, denoted as $inner_i$. Hence, we have established a unique one-to-one mapping between $(block_i, inner_i)$ and $p_i$. Through this encoding method, we efficiently reduce the dimensionality of the POI vocabulary from 100M to 400K, making softmax operations on the vocabulary range computationally efficient. Besides, when predicting the next POI, this hierarchical index allows for initially considering the blocks of geographic interest to the user and further predicting which specific POI within the blocks may interest the user, aligning more closely with the recommendation logic in POI scenarios. Compared with other hierarchical POI indexing strategy \cite{huang2022ernie}, our method uses fewer tokens to represent a POI, making the decoding process more efficiently.

Based on the hierarchical POI index, we propose a \textbf{spatiotemporal encoding module} for input information encoding. Each user action $s_i$ is represented using four tokens: ($u_i, block_i,$ $inner_i, a_i$). Here, $block_i$ and $inner_i$ are the aforementioned hierarchical POI indexes. The token $u_i$ represents the user's spatiotemporal context, including time $t_i$ and the user's geographic location $g^u_i$. $a_i$ denotes the action type. To further enrich the representation of POI information, we add the POI category information $c_i$ and geographic location $g^p_i$ onto the embedding of the $inner_i$ token. All time-related features are discretized along four dimensions: month, weekday, day, and hour; geographic features are discretized through geo-activated hashing functions. Each type of feature has its own embedding layer to map discrete features into continuous features. The input feature computation is formulated as follows:

\begin{align*}
&E(u_i) = w_1 * Emb_t(t_i) + w_2 * Emb_g(g^u_i) \\
&E(block_i) = Emb_p(block_i) \\
&E(inner_i) = w_3 * Emb_p(inner_i) + w_4 * Emb_c(c_i) + w_5 * Emb_g(g^p_i) \\
&E(a_i) = Emb_a(a_i)
\end{align*}

$Emb_t$, $Emb_g$, $Emb_p$, $Emb_c$, $Emb_a$ are the embedding layers for different types of features. $w_1$, $w_2$, $w_3$, $w_4$, $w_5$ are normalized dynamic weights for different types of embeddings. Compared with existing spacetime encoding methods \cite{luo2021stan,wu2020personalized} where time and geography information is encoded as side information, the key feature of this encoding module is the integration of users' spatiotemporal information as tokens within the sequence. First, this approach enables the model to leverage users' spatiotemporal data when predicting POIs, thus increasing the model's sensitivity to such information. Second, the spatiotemporal information can more effectively interact with other data in subsequent attention modules, enhancing the model's ability to utilize spatiotemporal information. 

In addition to the user action sequence, user profile data $up$, including gender, age, and occupational information, is formatted as text and concatenated at the beginning of the user sequence. After multiple layers of encoding, the output dimension of the model is projected to the vocabulary size dimension through a language modeling head, followed by a softmax function calculation to compute the probability of predicting the next token at each position, given by:

\begin{equation*}
P = \text{Softmax}(W_v * \text{LlamaDecoder}(\text{concat}(up, S)))
\end{equation*}


$W_v$ is the projection matrix. Finally, we employ cross-entropy loss as the loss function for pre-training. It is noteworthy that, when calculating loss, we only consider tokens corresponding to the block ID and inner ID in actions where $It=1$. Tokens at other positions are only used as input and do not contribute to the loss. The specific loss function is as follows:

\begin{equation*}
\begin{aligned}
\mathcal{L}_{pretrain} = -\sum\limits_{i=1}^{n-1}& It_{i+1} * (\log P(block_{i+1}|up, s_1, ..., s_i, u_{i+1}) + \\
& \log P(inner_{i+1}|up, s_1, ..., s_i, u_{i+1}, block_{i+1}))
\end{aligned}
\end{equation*}

\paragraph{\textbf{Training Strategy}} In the pre-training stage, we account for the potential complexity of patterns in user action sequences, especially in scenarios where users travel to different cities. Such actions may differ significantly from local actions in map applications, resulting in complex sequences that are challenging to learn from. Therefore, we draw on the strategy of curriculum learning \cite{soviany2022curriculum}, which involves first learning from simpler samples before gradually introducing more complex ones. Specifically, we divide all actions in $D$ based on the geographic location of both users and POIs into three statuses: local actions, pre-travel actions, and in-travel actions. Subsequently, we segment these sequences into multiple subsequences $S_1, S_2, ..., S_k$, where each subsequence contains actions from only one status. This results in a single-pattern dataset $D^{single}$, representing data under a singular pattern. In addition, sequences with multiple action statuses from $D$ consist of multi-pattern dataset $D^{multi}$. The pre-training process begins with training on the simpler dataset $D^{single}$, focusing on action prediction under specific patterns. It then progresses to training on more complex dataset $D^{multi}$, enhancing the model’s capability to predict transitions between user action patterns.

\subsubsection{Supervised Finetuning Stage}

As illustrated in Figure \ref{fig: Model} (b), after obtaining the pre-trained model, we proceed to use post-training strategies to optimize the model for downstream tasks. A key approach employed is Supervised Fine-Tuning (SFT). We classify SFT approaches into two types based on the output formats: embedding-based ranking SFT and generative ranking SFT.

\paragraph{\textbf{Data Processing}} The SFT stage focuses on the online recommendation scenario, where users are recommended $K$ POIs and choose to click on some of them based on their interests. Accordingly, the SFT data includes user profile $up$, action sequence $S$, spatiotemporal information at the time of request $u_{req}$, the set of clicked POIs as positive samples ($y=1$), and the set of exposed but unclicked POIs as negative samples ($y=0$).

\paragraph{\textbf{Embedding-Based Ranking SFT}} A key capability of pre-trained models is their encoding ability, which allows input data to be encoded into vectors with rich semantic information. Naturally, we attempt to finetune Spacetime-GR for producing high-quality embeddings. Referring to the dual-tower model frequently used in recommendation scenarios \cite{yang2020mixed,yi2019sampling}, we design a dual-tower structured fine-tuning task. As shown in Figure \ref{fig: Model} (b), the user-side features, including $up$, $S$, and $u_{req}$, are concatenated and fed into Spacetime-GR without the language modeling head. The output at the last position is extracted as the hidden state for the user side ($H_u$). The target POI hierarchical indexes ($block_p$, $inner_p$) are concatenated into a sequence of length 2 with associated POI side information. This input similarly undergoes Spacetime-GR encoding, yielding the hidden state for the POI side ($H_p$). Ultimately, these two hidden states are projected into low-dimensional vectors, serving as the user and POI embedding ($E_u, E_p$). The calculation is formulated as:

\begin{align*}
&H_u = \text{Spacetime-GR}(\text{concat}(up, S, u_{req}))[-1] \\
&H_p = \text{Spacetime-GR}(\text{concat}(block_p, inner_p))[-1] \\
&E_u = W_u * H_u \\
&E_p = W_p * H_p
\end{align*}

Subsequently, we calculate the cosine similarity between these two embeddings, using it to fit the sample's label. The training loss is based on the InfoNCE loss used in contrastive learning \cite{oord2018representation}, aiming to ensure that the similarity between the positive sample and the user embedding is higher than that between the negative sample and the user embedding. The formula is as follows:

\begin{equation*}
\begin{aligned}
&\mathcal{L}_{emb-sft} = \\
&-\sum\limits_i{\log\frac{\sum_j \exp(\cos(E_u^i, E_p^{i,j,+})/\tau)}{\sum_j \exp(\cos(E_u^i, E_p^{i,j,+})/\tau) + \sum_k \exp(\cos(E_u^i, E_p^{i,k,-})/\tau)}}
\end{aligned}
\end{equation*}

where $E_u^i$ stands for the user embedding of the $i$th sample, $E_p^{i,j,+}$ stands for the $j$th positive POI embedding in the $i$th sample, and $E_p^{i,k,-}$ stands for the $k$th negative POI embedding in the $i$th sample. $\tau$ is the temperature hyperparameter. The output embeddings could be served as features for downstream ranking models.

\paragraph{\textbf{Generative Ranking SFT}} The advantage of embedding-based ranking SFT lies in its independence of calculating user and POI embeddings, resulting in lower online computational cost and availability for offline inference. However, since there is no interaction between user and POI at the lower level of the model, it becomes challenging to leverage the potential of SpaceTime-GR to achieve more accurate recommendation results. Therefore, we design another fine-tuning task that directly outputs ranking scores based on user information and POIs. As illustrated in Figure \ref{fig: Model} (b), we concatenate the candidate POI set in the sequence format after the user information. The information for each POI is consistent with that used in embedding-based ranking SFT. To ensure that the scoring is not affected by the order of the candidate POIs, we modify the attention mask such that when calculating tokens corresponding to candidate POIs, only user information and the respective POI information are visible. We also ensure the position embeddings across different POIs to be the same. Finally, after model encoding, we extract the hidden representation corresponding to the inner ID token for each POI, named $H(inner_i)$, transform it through a classification head $W_c$ to obtain a probability $P_i$, and conduct binary classification based on cross-entropy loss. The calculation process is as follows:

\begin{align*}
&H = \text{Spacetime-GR}(\text{concat}(up, S, u_{req}, block_1, ..., inner_k))) \\
&P_i = \sigma (W_c * H(inner_i)) \\
&\mathcal{L}_{generative-sft} = -\sum\limits_i{y_i * \log P_i + (1- y_i) * log(1 - P_i)}
\end{align*}

The output probabilities could be directly used as ranking scores for downstream POI recommendation. 

\paragraph{\textbf{Multimodal POI Embeddings}} In the SFT stage, the model aims to enhance its ability to predict users' POI preferences. Thus, integrating additional dimensions of POI information is evidently beneficial for the task. In the pre-training stage, we only use the category and geographical location of POIs, which lacks textual and visual modalities. Therefore, at this stage, we attempt to leverage the multimodal capabilities of LLMs to enhance the input information on the POI side.\footnote{We do not incorporate the multimodal POI embeddings in the pre-training stage due to the increasing computational cost.} Specifically, as illustrated on the left side of Figure \ref{fig: Model} (b), we gather POI text information, including names, addresses, tags and reviews from map applications and concatenate them into a textual sequence. Additionally, we collect relevant images of POI as visual information. These two modalities are processed using a pre-trained multimodal LLM \cite{wang2024qwen2vlenhancingvisionlanguagemodels}. We perform average pooling over the hidden states of both modalities separately, finally aggregating them to obtain a POI embedding containing multimodal information. In practice, this information serves as the POI side information and is added to the inner ID token embedding, influencing both embedding-based and generative ranking SFT approaches.

\subsubsection{Alignment Stage}
Apart from generating embeddings and scores for ranking, we also aim to directly generate recommended POIs. In the pre-training stage, the model learns to predict the next POI in the user sequence. However, its ability of ranking remains limited. Thus, we employ post-training strategies to enhance the model's capability in joint recall and ranking.

\paragraph{\textbf{DPO Training}} Direct Preference Optimization (DPO) \cite{rafailov2023direct} is a prevalent post-training method characterized by learning from preference data. It involves calculating the model's output probabilities for positive and negative samples, thereby guiding the model to align with the distribution of preferred samples. Conveniently, in recommendation scenarios, such preference data naturally exists. User click samples can be treated as positive samples, whereas exposed non-clicked samples can be treated as negative samples. Therefore, we naturally apply the DPO framework in the alignment stage. We utilize the same training data as in the SFT stage, where each user action sequence has associated positive samples and negative samples. As illustrated in Figure \ref{fig: Model} (c), we concatenate the positive and negative POI candidates to the user action sequence in the same manner as in generative ranking SFT. After encoding by Spacetime-GR, we obtain the probabilities for block ID and inner ID for each candidate POI, with their product as the overall probability of the POI. Subsequently, we compute the DPO loss to optimize the model and the calculation formula is as follows:

\begin{equation*}
\begin{aligned}
&L_{DPO} = -\sum\limits_{i}\sum\limits_{j,k}(\log\sigma(\beta \log\frac{Align(p^{i,j,+})}{Ref(p^{i,j,+})} - \beta\log\frac{Align(p^{i,k,-})}{Ref(p^{i,k,-})}))
\end{aligned}
\end{equation*}

where $i$ represents the $i$-th sample, $p^{i,j,+}$ and $p^{i,k,-}$ represent the predicted probability for the $j$-th positive POI and the $k$-th negative POI in the $i$-th sample. $Align$ and $Ref$ represent the aligned model and the reference model. The former is the model being optimized and the latter is the base model for probability comparison. Both are initialized from the pre-trained model and only the aligned model is optimized while the reference model is frozen.

\paragraph{\textbf{Spatiotemporal Sensitivity}} The aforementioned alignment strategy could not only enhance general ranking capabilities but also improve the model's customization abilities. For instance, if we aim to make the model more sensitive to spatiotemporal information, we only need to construct targeted preference samples. Specifically, if we want to recommend food-related POIs during meal times, the positive sample would consist of spatiotemporal information corresponding to meal times, with the target POI being related to food. Conversely, the negative sample would involve the same spatiotemporal context but with a non-food-related POI as the target. Following the above DPO loss function, we can align the model's recommendation preferences with our requirements. Furthermore, by generalizing this capability, the framework can support alignment under any given preference based on the application scenario.

\section{Experiments}

In this section, we present the experimental details and results of Spacetime-GR on industrial datasets and public datasets.

\subsection{Experiments on Industrial Datasets}

\subsubsection{Dataset} 
We present the statistics of the industrial dataset in Table \ref{tab: experiment data}. The data used in pre-training stage follows the data cleansing strategy described in Section \ref{subsection: Pretraining Stage}. The data used in SFT and alignment stages comes from the downstream recommendation ranking data. Its average sequence length is greater than that of the pre-training data since users who use the recommendation function is more active than average, thus having more actions. 

\begin{table}[htbp]
\caption{\label{tab: experiment data} Industrial dataset information.}
\centering
\resizebox{0.47\textwidth}{!}{
\begin{tabular}{l|l|ccc}
\hline
Stage                             & Data Type  & Samples & Length & Candidate POI Num \\ \hline
\multirow{3}{*}{Pre-training}     & Train      & 578M    & 146.3            & -              \\
                                  & Validation & 19,794  & 142.6            & -              \\
                                  & Test       & 19,837  & 143.2            & -              \\ \hline
\multirow{3}{*}{SFT \& Alignment} & Train      & 31M        &  301.0           & 11.3               \\
                                  & Validation & 611K        &  334.5      &  10.5              \\
                                  & Test       & 553K        &  346.1           &  10.6              \\ \hline
\end{tabular}
}
\end{table}

\subsubsection{Experiment Settings}

We adopt the Llama 2 model architecture with 12 layer transformer blocks, an embedding dimension of 768, 32 attention heads, and an intermediate hidden size of 2048. Time embedding is the concatenation of embeddings for month, day, weekday and hour, where each has a dimension of 192. Geo embedding is formed by concatenating embeddings from 12 geo-activated hashing indexes, each with a dimension of 64. In the pre-training stage, the model is trained for 1 epoch, using the Adam optimizer. The initial learning rate is 1e-3, with 250 steps warmup and cosine decay. The minimum learning rate is 1e-4. Each training update step processes a total of 12,288 samples. 

Afterwards, we perform SFT and alignment separately based on the pre-trained model. For embedding-based ranking SFT and generative ranking SFT, we train on the data for 2 epochs. The initial learning rate is 1e-4, with 250 steps warmup and cosine decay. The minimum learning rate is 1e-5. Each update step processes a total of 1,440 samples. The temperature parameter $\tau$ in embedding-based ranking SFT is 0.1. For DPO alignment, the downstream ranking data is trained for 1 epoch. The initial learning rate is 1e-5, with 100 steps warmup. The minimum learning rate is 1e-6. Each update step processes 6,144 samples. The scaling parameter $\beta$ in DPO loss is 1.

\subsubsection{Evaluation Metrics}

We evaluate the results of different fine-tuned models based on their application scenarios. Embedding-based ranking SFT and generative ranking SFT produce embeddings and ranking scores that are used as features for POI ranking. Thus, we use the AUC score and CTR/CVR improvements to evaluate their performance compared with the online ranking model. The alignment stage optimizes Spacetime-GR to generate POIs, thus we directly compare the POI recommendation quality of Spacetime-GR and the online recommendation system by LLMs and humans.

\subsubsection{Results of SFT stage}

We first present the results of SFT models on POI ranking tasks. Given a user's request, for embedding-based ranking SFT model, we calculate the user and POI embeddings, normalize them and use them as input features for the online ranking model. As for generative ranking SFT model, we obtain the ranking scores for each candidate POI and merge these scores with those from the online ranking model. The combined score serves as the final ranking score. We present the AUC results of different applications in Table \ref{tab: SFT AUC Result}.

\begin{table}[htbp]
\caption{\label{tab: SFT AUC Result} AUC results on the POI ranking task.}
\centering
\begin{tabular}{l|r}
\hline
Methods                                          & AUC    \\ \hline
online ranking model &  0.7043 \\ 
+ embedding-based ranking SFT          &  0.7229\\
+ generative ranking SFT &  0.7272\\ 
+ embedding-based ranking \& generative ranking SFT                 & 0.7385 \\ \hline
\end{tabular}
\end{table}

After incorporating the features provided by SFT models, the AUC score obtains significant improvements—1.86 percentage points (pp) from embedding-based ranking model and 2.29pp from generative ranking model, proving that Spacetime-GR could import additional information gain on existing online ranking models. Moreover, combining both SFT models' outputs could further improve the performance, showing the complementarity of two strategies. 

Finally, we allocate 20\% of the traffic to the ranking model enhanced with SFT model outputs, and observe its performance for seven days. On average, this approach results in \textbf{6\%} improvements in CTR and \textbf{4.2\%} improvements in CVR for POIs overall, with the hypothesis test yielding a p-value approximately equal to zero under the one-sided t-test. This result further proves the applicability of Spacetime-GR in industrial online recommendation systems.

\subsubsection{Results of Alignment stage}

Next, we present the results of Spacetime-GR after alignment. To evaluate the actual effectiveness of POI recommendations, we design an assessment framework based on LLMs. Specifically, we randomly sample 800 requests from online users and retrieve the top-10 POIs recommended by both the online system and the aligned Spacetime-GR. We then construct a prompt consisting of the user's profile information, historical action sequence, and detailed descriptions of the top-10 POIs from each system. This prompt is fed to LLMs (e.g., GPT-4o, Qwen-Plus), which are instructed to assess the recommendation quality from two perspectives: system-level and POI-level. System-level evaluation involves a holistic comparison to determine which system's recommendations are superior overall, while POI-level evaluation involves assessing whether each recommended POI aligns with the user's interests. Then, we calculate which system recommends a larger number of POIs that correspond to the user's interests. In addition, we conduct a human evaluation involving seven volunteers who independently review the recommendation results and vote on which model performs better. The results in Table \ref{tab: Alignment LLM Result} indicate that our aligned Spacetime-GR is superior to the online model on both system level and POI level from LLM assessment and human evaluation as well.

\begin{table}[htbp]
\caption{\label{tab: Alignment LLM Result} Evaluation results from LLMs on POI recommendation.}
\centering
\begin{tabular}{l|rrr}
\hline
Spacetime-GR vs online model & Win    & Even   & Lose   \\ \hline
system level                    & 67.0\% & 2.0\%  & 31.0\% \\
POI level                       & 69.9\% & 10.7\% & 19.4\% \\ 
human                       & 55.2\% & 14.3\% & 30.5\% \\ \hline
\end{tabular}
\end{table}

\subsubsection{Computational Cost Analysis}

Spacetime-GR is trained for 7 days using 96 H20 GPUs during the pre-training phase, and 2 days on 16 H20 GPUs during the SFT and alignment phases. The computational cost of is relatively modest considering the scale of training on over hundreds of millions of data instances. Moreover, sufficient training is essential for the model to effectively capture information from such a large-scale POI corpus. In the online application, the SFT model generates representation and ranking results with an average latency of 10–20 milliseconds on T4 GPUs, demonstrating its ability to meet the low-latency requirements of real-time recommendation systems.

\subsection{Experiments on Public Datasets}

\subsubsection{Datasets \& Baselines}

To evaluate the generalizability of Spacetime-GR, we conduct experiments on three widely used public datasets, including Foursquare-NYC \cite{yang2014modeling}, Foursquare-TKY \cite{yang2014modeling} and Gowalla-CA \cite{cho2011friendship}. Note that all these public datasets consist of offline check-in records, which is different with the online POI recommendation scenario that Spacetime-GR focuses on. Thus, this experiment is conducted to test whether Spacetime-GR is available on different scenarios. The scale of users and POIs in all three datasets is on the order of thousands—much smaller than that of the industrial dataset. All the data preprocessing steps are kept the same with \cite{yan2023spatio}. We choose several generative recommendation baselines\footnote{Methods using LLMs are not compared due to injecting external knowledge and higher computational cost.}, including LSTM-based \cite{hochreiter1997long, zhao2020go, wu2020personalized}, attention-based \cite{luo2021stan} and transformer-based \cite{yang2022getnext,yan2023spatio} methods. We use top-1 hit rate accuracy to evaluate these methods.

\subsubsection{Experiment Settings}
It is important to note that all these public datasets consist of offline check-in records, thus the geographic information of users and POIs is the same for each record, and the action type is also uniform for all records. Thus, we adapt Spacetime-GR by removing the user geographic information and action type token. The hierarchical POI index also degrades into a single-level index. We only conduct the pre-training stage as the entire training process.
To reduce the effectiveness of model and computational complexity, we simplify spacetime-GR by setting the layer size as 2, making it comparable with the baseline methods. The initial learning rate is 5e-5; the minimum learning rate is 5e-6; the epoch number is 20; the batch size as 2. All the parameters are kept the same or three datasets. Other parameters are kept the same as those in the industrial dataset experiments. 

\begin{table}[htbp]
\caption{\label{tab: public dataset} Top-1 Accuracy on three public POI recommendation datasets.}
\begin{tabular}{l|rrr}
\hline
              & NYC   & TKY    & CA     \\ \hline
LSTM \cite{hochreiter1997long}         & 0.1305 & 0.1335 & 0.0665 \\
STGCN \cite{zhao2020go}         & 0.1799                     & 0.1716                     & 0.0961                     \\
PLSPL \cite{wu2020personalized}        & 0.1917                     & 0.1889                     & 0.1072                     \\
STAN \cite{luo2021stan}       & 0.2231                     & 0.1963                     & 0.1104                     \\
GETNext \cite{yang2022getnext}     & 0.2435                     & 0.2254                     & 0.1357                     \\
STHGCN  \cite{yan2023spatio}     & 0.2734 & 0.2950 & 0.1730 \\ \hline
Spacetime-GR & 0.2920                     & 0.2610                     & 0.1659                     \\ \hline
\end{tabular}
\end{table}

\subsubsection{Results}
The experimental results in Table \ref{tab: public dataset} have shown that our method outperforms traditional sequential recommendation methods and is comparable with STHGCN (superior on NYC but inferior on TKY and CA). Notably, STHGCN leverages other users’ sequences for prediction, whereas our method relies solely on the current user’s sequence and still achieves promising results. 

\section{Analysis}

\subsection{Ablation Study}

To further analyze the performance of Spacetime-GR, we conduct several ablation studies to validate the effectiveness of each module.

\subsubsection{Ablation Study of the Pre-training Stage}

In the pre-training stage, we focus on the effectiveness of the Spacetime-GR architecture and its training strategy. We adopt the hit rate metric to assess model performance. For each $s_i$ in user sequence $S$, the model is tasked with predicting the top-k POI set $S^{pred}_{i,k}$ given $\{s_1, ..., s_{i-1}\}$. The hit rate @k is computed as 

$$hr@k = \frac{\sum\limits_i\mathbb{I}(s_i \in S^{pred}_{i,k})}{|S|}$$

We present hit rate results at $k$=1 and 100. When decoding the next POI, we adopt the beam search strategy by first predicting 10 block IDs, and predict 10 inner IDs for each block ID. At last, we calculate the probability of the 100 candidates and select the top-k POIs with highest probabilities as the top-k POI set. Table \ref{tab: Pretraining Result} presents the results of different model structures. \textbf{GPT-based} represents a baseline method where we choose the 12-layer decoder-only transformer as backbone and train to generate the next token given previous user actions. \textbf{Spacetime-GR} represents our proposed method. \textbf{w/o spatiotemporal info} represents Spacetime-GR without incorporating time or geographic information. \textbf{w/o hierarchical POI index} represents Spacetime-GR with traditional hashing index instead of geographic-aware hierarchical index. \textbf{w/o curriculum learning} represents training Spacetime-GR with all the data together, not considering different patterns under the curriculum learning strategy.

\begin{table}[htbp]
\caption{\label{tab: Pretraining Result} Hit rate results of pre-training.}
\centering
\resizebox{0.4\textwidth}{!}{
\begin{tabular}{l|rrrr}
\hline
Methods                                 & \multicolumn{1}{l}{hr@1}     & \multicolumn{1}{l}{hr@100} \\ \hline
GPT-based                       & 0.0688                               & 0.2195                     \\ \hline
Spacetime-GR                   & \textbf{0.1525}                  & \textbf{0.4721}                     \\ 
w/o spatiotemporal info                      & 0.1007  & 0.3671 \\
w/o hierarchical POI index & 0.1328                                     & 0.3480                     \\
w/o curriculum learning                 & 0.1463                               & 0.4624                     \\ \hline

\end{tabular}}
\end{table}

From the results, we can conclude that our model structure outperforms the traditional GPT model on all the hit rate metrics, benefit from the more advanced Llama structure and our spatiotemporal encoding strategy. Focusing on separate components of Spacetime-GR, what affects most is the spatiotemporal information. Removing this component leads to a 5pp drop on hit rate @1 and 10.5pp on hit rate @100. It proves the importance of spatiotemporal context on this task. Another key design is the geographic-aware hierarchical POI index. It is better than random hashing index with 2pp on hit rate @1 and shows greater gains on hit rate @100, which could be attributed to the ability to generate higher-quality candidate block IDs more easily. Additionally, curriculum learning also shows improvements on naive training strategy by 1pp on hit rate @100. All these settings help enhance the model’s ability to predict the next POI.

\subsubsection{Ablation Study of the SFT Stage}

In the SFT stage, we focus on comparing the performance of two different SFT strategies and the effectiveness of multimodal POI embeddings. Table \ref{tab: SFT offline Result} shows the performance of different SFT methods. First, comparing embedding-based ranking SFT and generative ranking SFT, we can conclude that based on deep interactions between user sequence and candidate POIs, generative ranking SFT can better exploit the capability of the pre-trained model. However, it also introduces higher computational complexity as the number of candidate POIs increases. Next, training from scratch performs much worse than from pre-trained model, highlighting the importance of the pre-training stage. Finally, adding multimodal embeddings could obtain significant improvements (1.3pp for embedding-based ranking SFT and 0.5pp for generative ranking SFT), proving the information gain from multimodal LLM embeddings.

\begin{table}[htbp]
\caption{\label{tab: SFT offline Result} AUC results in the SFT stage.}
\centering
\begin{tabular}{l|r}
\hline
Methods                                          & AUC    \\ \hline
embedding-based ranking SFT from scratch &  0.6621\\ 
embedding-based ranking SFT             &  0.7080\\
embedding-based ranking SFT + multimodal &  0.7214\\ \hline
generative ranking SFT from scratch                      & 0.6648 \\
generative ranking SFT                                   & 0.7371 \\ 
generative ranking SFT + multimodal & 0.7425 \\ \hline
\end{tabular}
\end{table}

\subsubsection{Ablation Study of the Alignment Stage}

The alignment stage mainly focuses on improving the model's joint recall and ranking ability, thus we adopt hit rate @1 and @10 as the evaluation metric to assess the model's joint recall and ranking ability compared with the pre-trained model. We choose the most recent 5 POI actions as target and the rest as input. A prediction is considered a hit if it matches any of the 5 target POIs.

\begin{table}[htbp]
\caption{\label{tab: Alignment hitrate Result} Hit rate results in the alignment stage.}
\centering
\begin{tabular}{l|rr}
\hline
Methods    & hr@1   & \multicolumn{1}{l}{hr@10} \\ \hline
pre-trained & 0.1960 & 0.4295                   \\
DPO        & 0.2006 & 0.4520                   \\
S-DPO      & 0.2008 & 0.4512                   \\ \hline
\end{tabular}
\end{table}

We present the results in Table \ref{tab: Alignment hitrate Result}. Compared with the pretrained model, alignment with DPO could improve the model's ranking ability in a large extent, with 2.25pp gain on hit rate @10, contributed by more accurate probability estimation through the DPO loss. Meanwhile, we also compare the S-DPO loss\cite{chen2024softmax} with traditional DPO loss, and the results show that two losses are comparable.




\subsection{Further Analysis}

In this section, we are going to answer the following questions.
\begin{itemize}
    \item RQ1: How does the model perform under different lengths of user action sequences?
    \item RQ2: What are the key advantages of our model's recommendations compared with the online system?
    \item RQ3: What factors contribute to the model's improvements in POI ranking?
\end{itemize}

\begin{figure}[htbp] 
  \centering
  \includegraphics[width=0.35\textwidth]{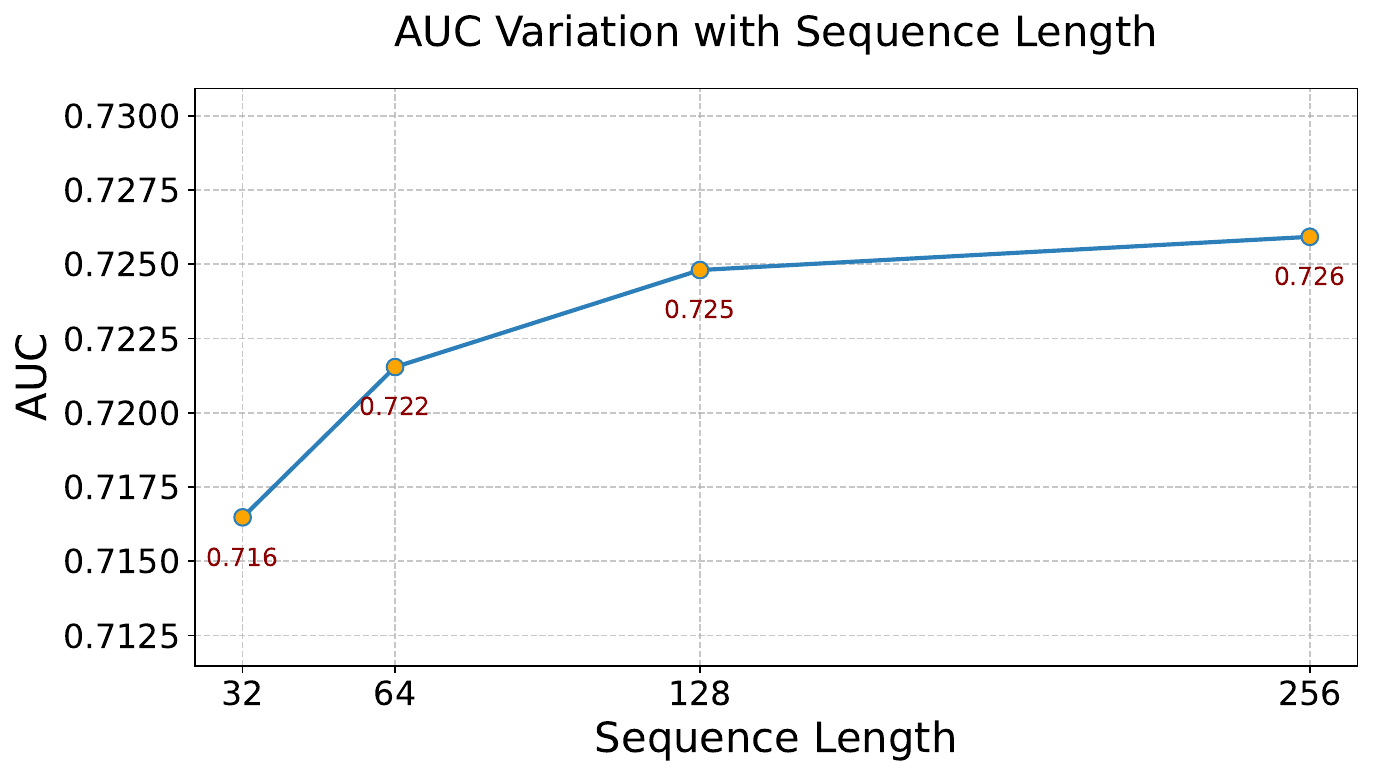}
  \caption{\label{Fig: length} SFT performance variation with sequence length.}
\end{figure}

\paragraph{\textbf{RQ1: Influence of Input Sequence Length}} In recommendation systems, sequence length has consistently been a critical hyperparameter. A shorter sequence length may limit the model's ability to fully capture user preferences, while a longer sequence length could substantially increase computational demands. Consequently, we investigate the impact of different sequence lengths on AUC scores in the generative ranking SFT stage. We restrict the upper length limit of input user action sequence to 32, 64, 128, and 256, and conduct SFT under each of these limits. Results\footnote{The results are not consistent with those presented in Table \ref{tab: SFT offline Result} since the test datasets are different.} in Figure \ref{Fig: length} have shown that the model's performance does indeed improve with increased sequence length, though the improvement becomes marginal beyond the length of 128. This indicates that the model primarily relies on recent interactions. Therefore, in the practical deployment, we set the maximum length to 128 to keep the balance between model effectiveness and computational efficiency.

\paragraph{\textbf{RQ2: The Discoverability of POI Recommendations}} To analyze the distribution differences between the results recommended by our model and those from the online system, we introduce a novel metric named \textbf{Discovery Rate}, which measures the novelty of the recommended POIs. We define $S_{rec}(k)$ as the set of top-$k$ recommended POIs, and $S_{new}(k, m)$ as the subset of top-$k$ recommended POIs not interacted by the user in the past $m$ active days. The definition is as follows:

$$D(k,m) = \frac{|S_{new}(k,m)|}{|S_{rec}(k)|}$$

\begin{figure}[htbp] 
  \centering
  \includegraphics[width=0.4\textwidth]{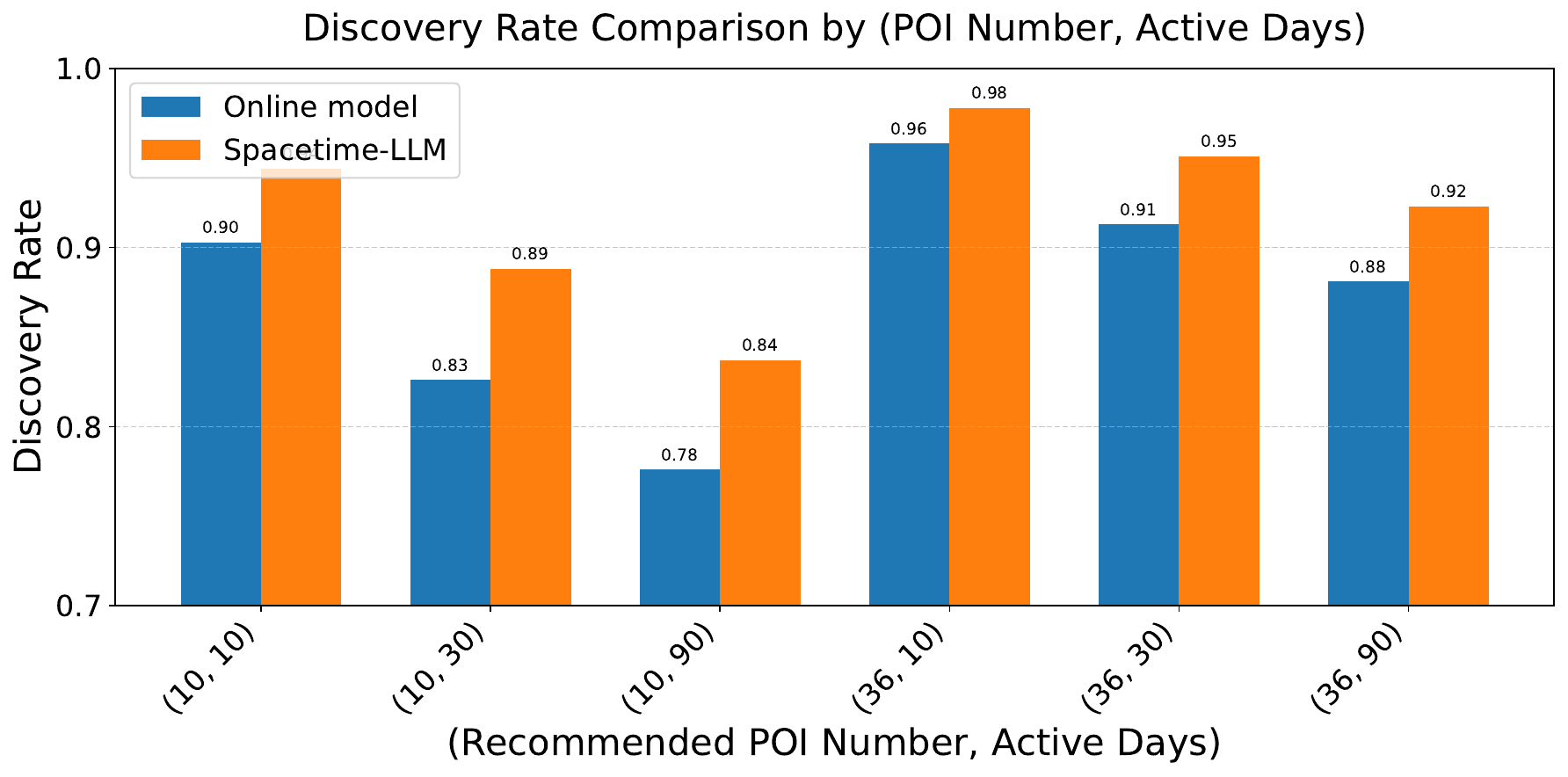}
  \caption{\label{Fig: Discovery} Discovery rate of POI recommendation.}
\end{figure}

Higher discovery rate values indicate that the model's recommendations are more novel. As shown in Figure \ref{Fig: Discovery}, our model consistently exhibits stronger discovery capability compared with the online baseline model across various parameter settings. From the perspective of POI number, increasing the number of recommended POIs enhances discovery rate, indicating that recommending more POIs reduces the overlap ratio with users’ historical interactions. From the perspective of active days, extending the time span results in reduced discovery values. Nevertheless, our model consistently outperforms the baseline model, highlighting the distinctiveness of the recommendation results from Spacetime-GR.

\begin{figure}[htbp] 
  \centering
  \includegraphics[width=0.4\textwidth]{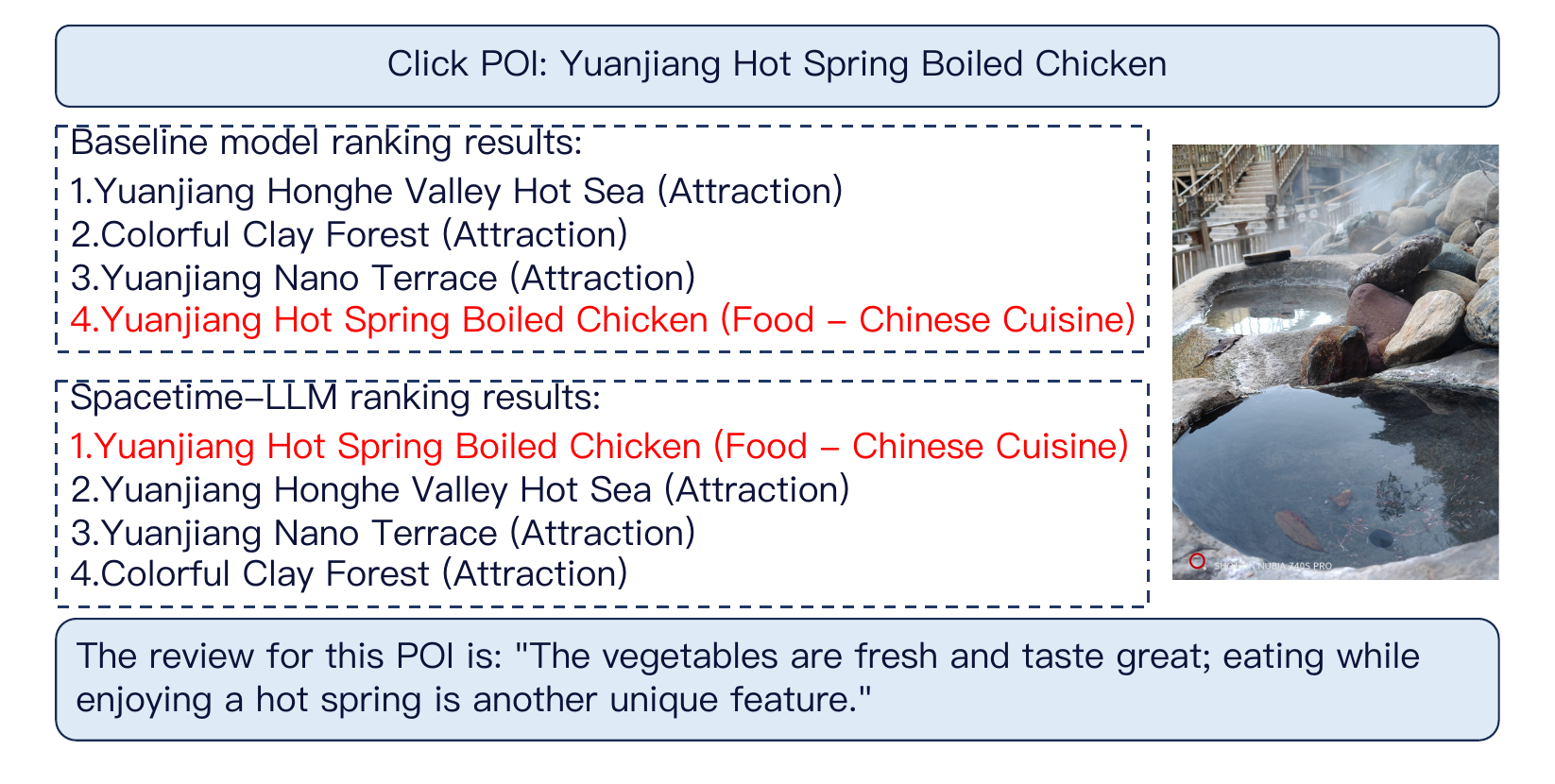}
  \caption{\label{Fig: Case Study} A case study on POI ranking.}
\end{figure}

\paragraph{\textbf{RQ3: Information Gain on POI Ranking}} To further analyze the improvement in our model's POI ranking performance, we randomly sample a case from the test set where our model outperforms the online system. As presented in Figure \ref{Fig: Case Study}, the user's historical actions reflect an interest in hot springs. Leveraging the capabilities of pre-training, our model successfully uncovers this interest preference. Moreover, during the SFT stage, we introduce the multimodal representation, with POI reviews mentioning related information on hot springs, further enhanced by the image. This enables the model to assign a higher score to the target POI, thereby improving the ranking effectiveness.

\section{Conclusion}

This work proposes Spacetime-GR, a spacetime-aware generative online POI recommendation model. It extends the sequence modeling ability of generative methods and adapts to POI recommendation scenarios through flexible spatiotemporal
information encoding. A complete training paradigm (pre-training and post-training) for POI recommendation is also invented to support various downstream application scenarios. Experimental results demonstrate its effectiveness on public datasets and large-scale POI recommendation tasks in industrial settings. Notably, the proposed paradigm can be adapted to other recommendation scenarios beyond POI recommendation. In future work, we aim to further enhance the generalization capabilities of our approach.

\section*{GenAI Usage Disclosure}

We use GPT-4o to polish the expression of the content and do some of the translation and grammar checking work. Besides, part of the icons in Figure \ref{fig: Intro}, including the cartoon person, food, computer and attraction pictures, are generated by GPT-4o for better illustration.

\bibliographystyle{ACM-Reference-Format}
\bibliography{sample-base}










\end{document}